# Near-Field Topology-Optimized Superchiral Metasurfaces for Enhanced Chiral Sensing

Zhongjun Jiang, Soyaib H Sohag, and You Zhou*

*Department of Physics and Optical Science, University of North Carolina at Charlotte, Charlotte, NC, USA, 28223*

*Email: yzhou33@charlotte.edu

**Abstract**

The detection and discrimination of molecular chirality are essential for advancing pharmaceutical and biological applications. While nanophotonic platforms offer a route to enhance chiral light–matter interactions, existing device concepts for chiral sensing remain heuristic, resulting in limited chiral enhancement and control over chiral hotspot placement within nanostructures. Here, we introduce an inverse-design framework that directly optimizes superchiral near fields in photonic nanostructures and demonstrate its powerful opportunities for enantioselective analysis. We first show that freeform achiral metasurfaces can be optimized to achieve an 800-fold chiral density enhancement, with fully customizable chiral hotspot placement for direct molecular interaction. We further demonstrate ultrasensitive detection of chiral analytes and achieve a 116-fold increase in detection sensitivity over the native enantiomeric response. Lastly, we extend our platform to determine chiral-molecule concentration and resolve enantiomeric excess in chiral mixtures. Our framework offers a generic route to enabling nanophotonic platforms for detecting chiral compounds and can be integrated with a broad range of spin-based photonic materials for applications in valleytronics, chiral emission control, and topological photonics.

Chirality is a geometric property of an object that cannot be superimposed on its mirror image. Such spatial asymmetry at the molecular scale gives rise to enantiomeric pairs that often exhibit distinct biological and pharmacological properties[1]. In a chiral mixture, the enantiomeric composition is typically identified by circular dichroism (CD)[2], based on the differential absorption of right- and left-circularly polarized light (RCP and LCP). However, the enantioselective sensitivity of conventional CD spectroscopy is inherently weak due to the large mismatch between molecular dimensions and the helical pitch of circularly polarized light (CPL). Rapid advances in nanotechnology have enabled nanophotonic platforms to resonantly enhance light–matter interactions[3–5], offering a route to enhanced enantioselective detection in a compact manner[6–10]. These developments have led to various nanophotonic concepts for applications spanning from far-field wavefront shaping[11–17] to functionalities enabled by enhanced near fields[18–24]. However, a design framework that effectively enhances optical chirality in photonic nanostructures remains elusive.

Achieving sensitive enantioselective detection requires photonic sensors that confine the helical twist of CPL to subwavelength dimensions, thereby generating superchiral near fields[25–28]. The optical chiral density is defined as[8,29]

$$C = -\frac{k_0}{2c_0}\text{Im}(\mathbf{E} \cdot \mathbf{H}^*) \qquad (1)$$

where $\mathbf{E}$ and $\mathbf{H}$ are the complex electric and magnetic field vectors, and $k_0$ and $c_0$ are the wavenumber and speed of light in free space, respectively. As made explicit by Eq. (1), achieving large superchiral fields requires the excitation of strong, localized $\mathbf{E}$ and $\mathbf{H}$ fields that are (i) spatially and spectrally overlapped, (ii) directionally aligned, and (iii) phase-shifted by $\pi/2$. To facilitate direct interaction with target analytes, the chiral hotspots should also form in the open regions of the nanostructures where the molecules reside. Furthermore, in contrast to chiral metasurface concepts that aim to enhance far-field CD[30–34], a chiral sensor should exhibit minimal chiroptical background to isolate the intrinsic analyte response[7,29,35,36], which necessitates geometrically achiral layouts with mirror or $C_n$ ($n > 2$) rotational symmetry. Achieving these criteria through nanostructure engineering is a highly nontrivial task due to the lack of precise analytical correlations between nanoscale geometries and near-field distributions. To date, the existing chiral sensing paradigm has relied on an empirical identification of simple geometries described by a small set of parameters, followed by fine-tuning these parameters through full-wave simulations[29,35–39]. While this approach has produced an alphabet of meta-atom templates that forms the foundation of chiral sensing research, it offers limited control over both the magnitude and the spatial placement of chiral hotspots. As a result, the chiral enhancements in these rationally designed devices remain modest and are predominantly confined within the nanostructures[29,35–39].

Here, we develop a novel inverse-design framework that explicitly engineers superchiral near fields in dielectric metasurfaces. While inverse design techniques have been widely used to shape the far-field responses of metasurfaces[40–45], our approach bridges intricate near-field engineering with freeform topology optimization by targeting near-field chiral figures-of-merit (FoM). The schematic workflow of our design concept is shown in Fig. 1. First, the spatial profile of the superchiral field is nucleated by defining the FoM as $\text{Im}(\mathbf{E} \cdot \mathbf{H}^*)$ at a probe point inside the metasurface (Fig.1 left), thereby directly linking maximization of this FoM to enhanced local chiral density. The nanostructure design space is constrained to be geometrically achiral, so the resulting far-field CD arises solely from the chiral analyte. The frequency, handedness, and spatial position of the FoM can be tailored at will, allowing chiral hotspots to be placed outside the metasurface structures for direct molecular interaction. To couple free-space waves to the desired chiral modes, we develop a near-field topology-optimization framework based on adjoint-variable method[46,47], in which a combination of far- and near-field excitation sources are utilized for forward and adjoint simulations[48] (Fig.1, middle). Our method explores a large freeform design space by rigorously accounting for the complex relationships between optical near fields and nanoscale geometry, thereby pushing the chiral-enhancement limits beyond empirical, template-based designs. Through multi-objective optimization, our framework naturally extends to devices hosting multifunctional chiral responses for broad spin-selective photonic applications[49–51]. Leveraging these capabilities, we realize freeform superchiral metasurfaces capable of ultrasensitive detection and discrimination of (R/S)-1,2-propanediol enantiomers (Fig. 1, right).

The metasurface consists of a subwavelength array of freeform meta-atoms arranged on a square lattice with a period of 720 nm (Fig. 2a). We choose silicon as the dielectric material due to its low loss, high refractive index, and ability to support Mie-type resonant modes[4,52,53]. A 460-nm thick silicon layer is selected to support excitation of multipolar resonances consisting of spectrally overlapped electric and magnetic modes. The FoM is defined as the pointwise quantity $\text{Im}(\mathbf{E} \cdot \mathbf{H}^*)$ that measures local chiral density and is maximized at the unit cell center under free-space CPL illumination. We enforce mirror symmetry on the freeform design space to impose geometric achirality (Fig. 2b, left), thereby suppressing structural chiroptical background. To ensure that the chiral hotspot can form outside the silicon nanostructures (Fig. 2b, right), we impose a small air void centered at the FoM probe point as the molecular host region.

We employ the adjoint solver in the open-source finite-difference time-domain package Meep[54], coupled with the Adam optimizer in PyTorch[55], to perform gradient-descent optimization (see optimization details Supplementary Section 1). The optimization trajectory showing FoM enhancement (log scale) and topology evolution is presented in Fig. 2c. Over the course of optimization, the FoM increases consistently and ultimately reaches more than an 800-fold enhancement compared to the starting point. The inset shows the topology transformation that evolves from a grayscale permittivity distribution to a binary freeform structure consisting of air and silicon. The top-view chiral density maps show a steady increase over the course of iteration, ultimately leading to strong chiral-field enhancement in the final binarized device. The top- and side-view overlays of the superchiral near field on the nanostructure (Fig. 2d) further confirm a localized chiral hotspot confined within the air gap. To uncover the resonant mechanisms behind the enhanced chiral field, we perform multipolar decomposition of the metasurface near fields from the current density distributions induced in the nanostructures using the open-source software MENP[56]. The multipolar decomposition as a function of wavelength (Fig. 2e) reveals the excitation of two dominant dipole modes in the form of spectrally overlapped in-plane electric $E_x$ and magnetic $H_x$ resonances. The relative phase retardation between the two modes, shown in Fig. 2f, is roughly $\pi/2$ at the designed wavelength, satisfying the condition for enhanced chiral density that scales as $\text{Im}(\mathbf{E} \cdot \mathbf{H}^*)$. It is worth noting that such geometrically smooth structure represents only one local optimum in a non-convex optimization landscape, which we selected to promote robust fabrication. Notably, our approach enables exploration of a substantially larger freeform achiral design space. As shown in Fig. 2g, we further optimize the same chiral FoM from a randomly initialized permittivity distribution and obtain a physically non-intuitive achiral geometry exhibiting similar chiral hotspot formation.

To experimentally validate the design, we fabricate the optimized metasurfaces on a 462-nm-thick amorphous-silicon film grown by plasma-enhanced chemical vapor deposition. The freeform nanostructures are defined using standard electron-beam lithography and reactive-ion etching (see fabrication details in Supplementary Section 2). A scanning electron microscope image of the fabricated device is presented in Fig. 3b and shows well-defined geometric features consistent with the design. We first evaluate the far-field response of the metasurface without a chiral overlayer (Fig. 3a). As shown in Fig. 3c, the simulated transmittance spectra reveals two dips corresponding to excitation of electric and magnetic dipole modes shown in Fig. 2e. The far-field transmittance spectra (Fig. 3a) are characterized under CPL illumination from a supercontinuum laser source, with the transmitted light collected by a customized imaging system and relayed to a spectrometer (see details of measurement setup in Supplementary Section 2). The measured transmittance shown in Fig. 3d exhibits good agreement in spectral line shape with the simulation.

We further validate the device's sensing performance by coating it with a thin chiral overlayer (Fig. 3e). The molecules are modeled as an isotropic chiral medium with refractive indices $n$ and Pasteur parameter $\kappa$. We select a weak chiral overlayer with $n = 1.34 - 0.001i$ and $\kappa = (7 - 1.5i) \cdot 10^{-3}$ uniformly coated on the metasurface. The resulting coupled electromagnetic response is obtained from full-wave simulations in COMSOL Multiphysics. We first compute the optical chiral density $C$ inside the nanostructures and evaluate the LCP-RCP difference, $\Delta C = |C_L| - |C_R|$, as a measure of metasurface-enhanced chiral sensitivity. As shown in Fig. 3f, the top view of the $\Delta C$ distribution exhibits a pronounced localized peak with more than a $10^3$-fold enhancement over the native molecular signal. We define CD as $(T_L - T_R)/(T_L + T_R)$ in this work, where $T_{L,R}$ denote the transmission under LCP/RCP. The simulated CD spectra (Fig. 4g) reveal a 116-fold increase in peak CD for the metasurface sensor compared to a bare substrate.

We then experimentally assess the sensing performance by coating the metasurface with a thin layer of (S)-(+)-1,2-propanediol solution (see Supplementary Section 2 for analyte preparation). The measured CD spectrum (Fig. 3h) shows more than 26× enhancement in CD contrast, with a line shape in good agreement with simulation. We attribute the reduced CD enhancement in the experiment to fabrication imperfections, non-uniform chiral overlayer and the simplified simulation model. Further improvements can be achieved by ensuring uniform analyte delivery and imposing more stringent feature-size constraints to reduce sensitivity to fabrication imperfections[44,57–60]. We note that while structural CD from the metasurface is, in principle, eliminated by its geometrically achiral layout, residual device chirality may still arise from oblique incidence due to slight substrate tilt or other measurement asymmetries. To assess these effects, we characterize the metasurface without chiral overlayers and observe negligible CD under the same measurement conditions (Supplementary section S3).

Lastly, we investigate enantioselectivity of the device as a function of concentration and demonstrate its ability to resolve enantiomeric purity for arbitrary chiral mixtures. We first perform full-wave simulations of the metasurfaces covered with an enantiopure chiral overlayer for different values of the Pasteur parameter. We use the scaled Pasteur parameter $\kappa/\kappa_0$ with $\kappa_0 = -(2.3 - 0.5i) \times 10^{-3}$; thus $\kappa/\kappa_0 = +1(-1)$ denotes 100% R- (S-) enantiomer. Fig. 4a shows the $\Delta C$ distributions inside the nanostructures for different $\kappa/\kappa_0$, revealing chiral contrast that grows with molecular concentration, consistent with the monotonic increase in the CD spectra (Fig. 4b). The extracted peak CD from each concentration (Fig. 4c) shows linear scaling with $\kappa/\kappa_0$, thereby providing a quantitative readout of enantiomeric concentration.

We further extend our platform to determine enantiomeric excess and purity in chiral molecular mixtures. This demonstration reflects practical pharmaceutical scenarios, where most drugs are prepared as mixtures of enantiomers, and the ability to accurately assess enantiomeric purity is crucial for drug synthesis and pharmaceutical development. The enantiomeric excess (*e.e.*) is defined as a measure of purity[61]:

$$e.e. = \frac{C_R \cdot V_R - C_L \cdot V_L}{C_R \cdot V_R + C_L \cdot V_R}$$

where $C_R$ and $C_L$ denote the concentrations of R- and L-handed chiral molecule solutions, respectively, and $V_{R,L}$ is the volume of R- (L-) handed molecules. We prepared mixtures of (R)- and (S)-1,2-propanediol at varying compositions while keeping the total volume fixed at 0.80 mL. The enantiomer fraction was varied over the full composition range, from purely L-enantiomer to purely R-enantiomer, with the complementary fraction provided by the opposite enantiomer. Each mixture was adsorbed onto the sensor to record the CD signal. Between measurements, the residual adsorbates were removed by sequentially rinsing the surface with dimethyl sulfoxide (DMSO), acetone, isopropyl alcohol (IPA) and de-ionized water. Fig. 4d shows the peak CD as a function of enantiomeric excess, revealing a linear relationship with the concentration imbalance between the two enantiomers. These results highlight the robustness of our approach for quantitative readout of enantiomeric concentration and purity. The sensing platform can be further improved at the system level by integrating the sensor with microfluidic flow cells[62] that allow uniform, actively controlled delivery of chiral analytes.

In summary, we introduce and demonstrate a near-field topology optimization framework for creating and tailoring superchiral near fields in freeform dielectric metasurfaces. Our approach directly optimizes local chirality density in photonic structures by rigorously accounting for the complex relationships between optical near-fields and nanoscale geometry. Compared with conventional template-based designs, our framework offers two distinct advantages for ultrasensitive enantioselective analysis. First, it explores a large freeform design space to push the chiral enhancement limits. Second, it enables designer control over chiral hotspots locations within the nanostructure to facilitate direct molecular interaction. In addition to enantioselective analysis, metasurfaces supporting enhanced chiral near fields can be integrated with wide range of spin-based photonic materials to advance emerging areas such as valleytronics[49–51], chiral emission control[63–65], and topological photonics[66]. Furthermore, our framework can be extended to devices hosting complex near-field modal profiles[48,67–71], thereby providing a generic route to resonant photonic platforms for applications spanning spontaneous-emission control[72–75], nonlinear optics[22,23,76,77], optomechanics[78,79], and photochemistry [80,81], where strong light-matter interactions are required.

## Data Availability

The data that support the plots within this paper and other findings of this study are available from the corresponding authors upon reasonable request.

## Acknowledgments

Y.Z. acknowledges support from the UNC Charlotte Faculty Research Grant, the Center for Metamaterials, and UNC Charlotte start-up funds and National Science Foundation (NSF) under awards ECCS-2501853.

## Contributions

Y.Z. and Z.J. developed the idea. Z.J. conducted the design, modeling, theoretical analysis and experimental measurements. S.S. fabricated the samples. Y.Z. and Z.J. wrote the manuscript with input from all authors. The project was supervised by Y.Z.

## Competing interest

The authors declare no competing interests.

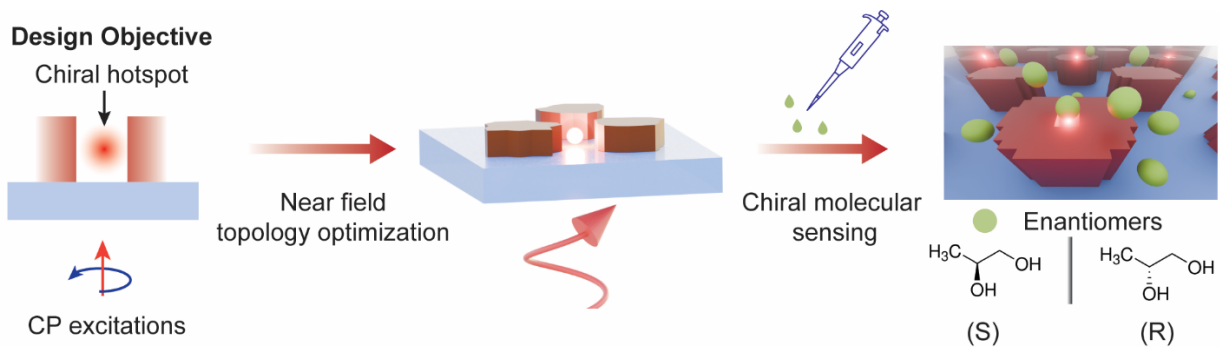

**Fig. 1 | Framework for engineering superchiral near fields in freeform metasurfaces.** Optical chiral density inside the photonic nanostructure is defined as the design objective and maximized under circularly polarized (CP) illumination via near-field topology optimization (left). The resulting superchiral hotspots are formed outside the nanostructures (middle), enabling ultrasensitive detection and discrimination of (R/S)-1,2-propanediol enantiomers (right).

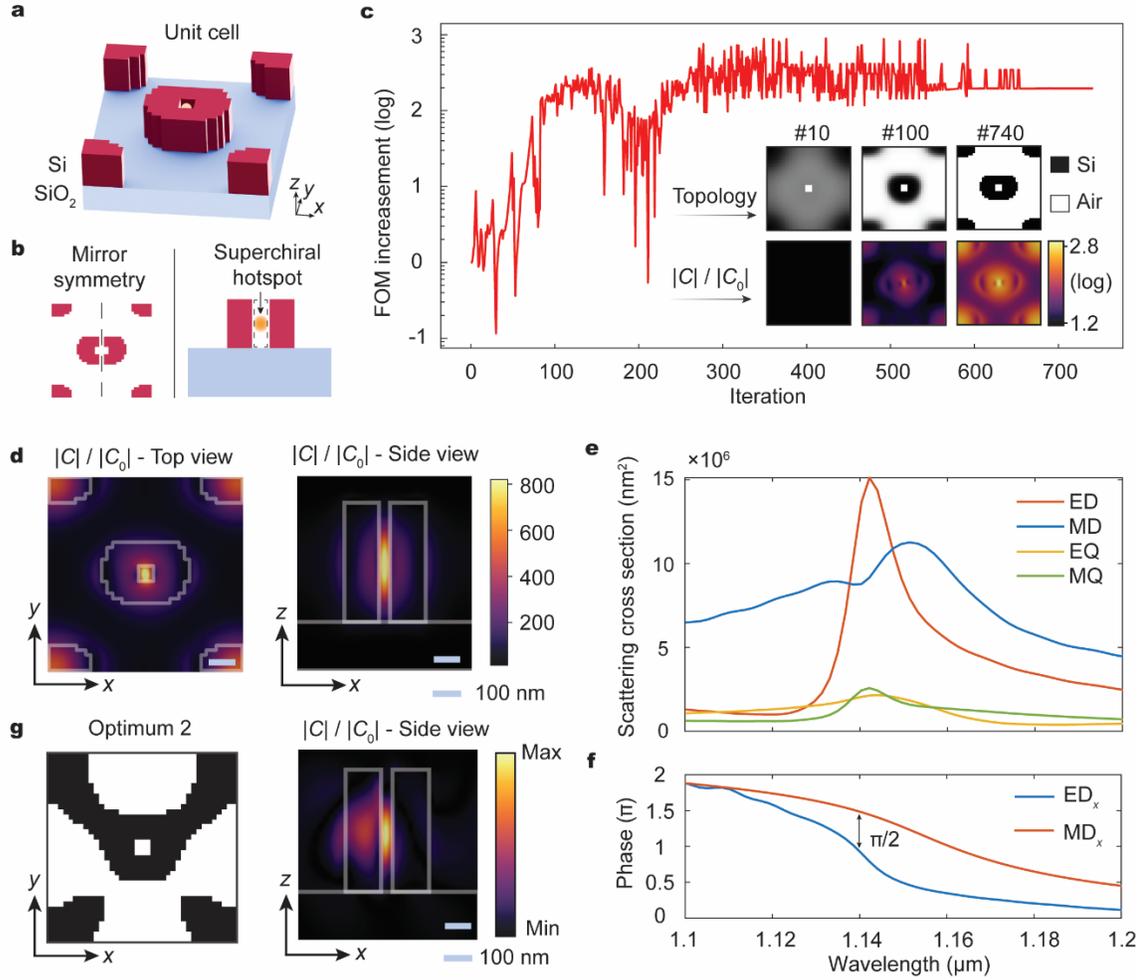

**Fig. 2 | Superchiral freeform metasurfaces.** (a) Unit cell of the optimized metasurface. (b) key design considerations. First, mirror symmetry constraint is applied to the design space to ensure a geometrical achiral layout (left). Second, a small air void is imposed and centered at the target hotspot to enable direct molecular access (right). (c) Optimization trajectory showing the enhancement of the Figure of Merit (FoM) during optimization. Insets: structural and normalized chiral density (log scale) evolution within the unit cell. (d) Top view (left) and side view (right) of chiral density distributions (linear scale) overlaid on the nanostructures, confirming a strongly localized chiral hotspot confined within the air gap. (e) Multipolar decomposition of the optimized metasurface. Abbreviations: ED (electric dipole), TD (toroidal dipole), MD (magnetic dipole), EQ (electric quadrupole), MQ (magnetic quadrupole). (f) Phase delays confirming a $\pi/2$ phase difference between the ED and MD resonances. (g) A second freeform achiral design obtained from a random initialization (top view) and its normalized chiral-density distribution (side view).

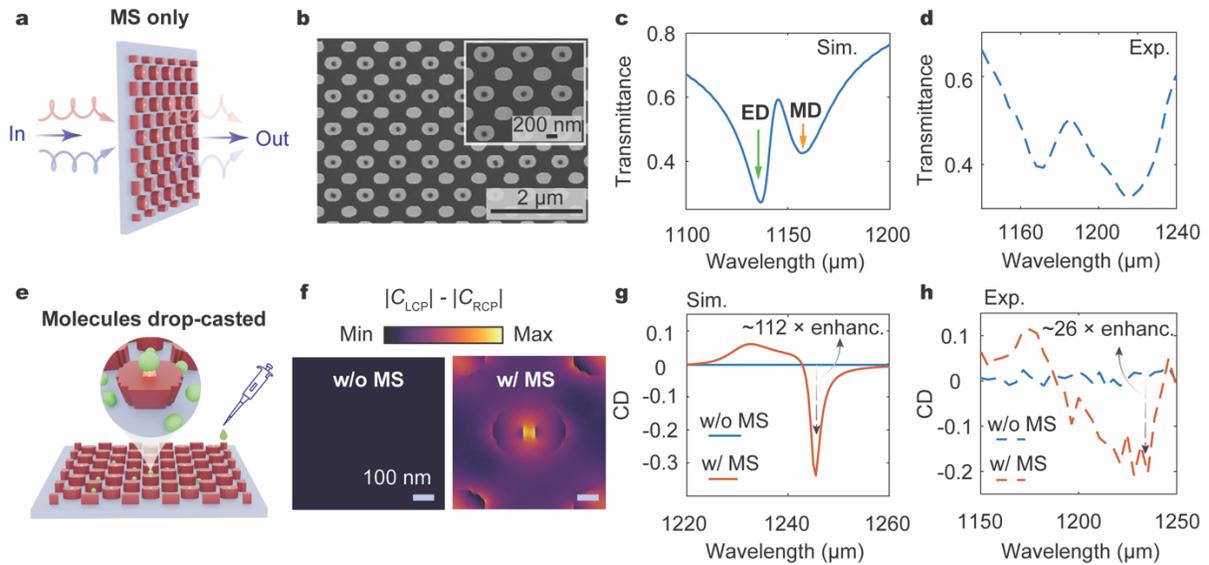

**Fig. 3 | Experimental demonstration of chiral sensing.** (a) Far-field device characterization. (b) Scanning electron microscope images of the fabricated metasurface sensor. (c-d) Simulated (c) and measured (d) transmittance spectra of the metasurface without chiral analyte. (e) Chiral sensing by drop-casting a chiral overlayer onto the metasurface. (f) Differential chiral density distribution $\Delta C = |C_L| - |C_R|$ for molecules without (left) and with (right) metasurface. (g-h) Simulated (g) and experimental (h) CD spectra for molecules on the metasurface (orange) and on a bare glass substrate (blue).

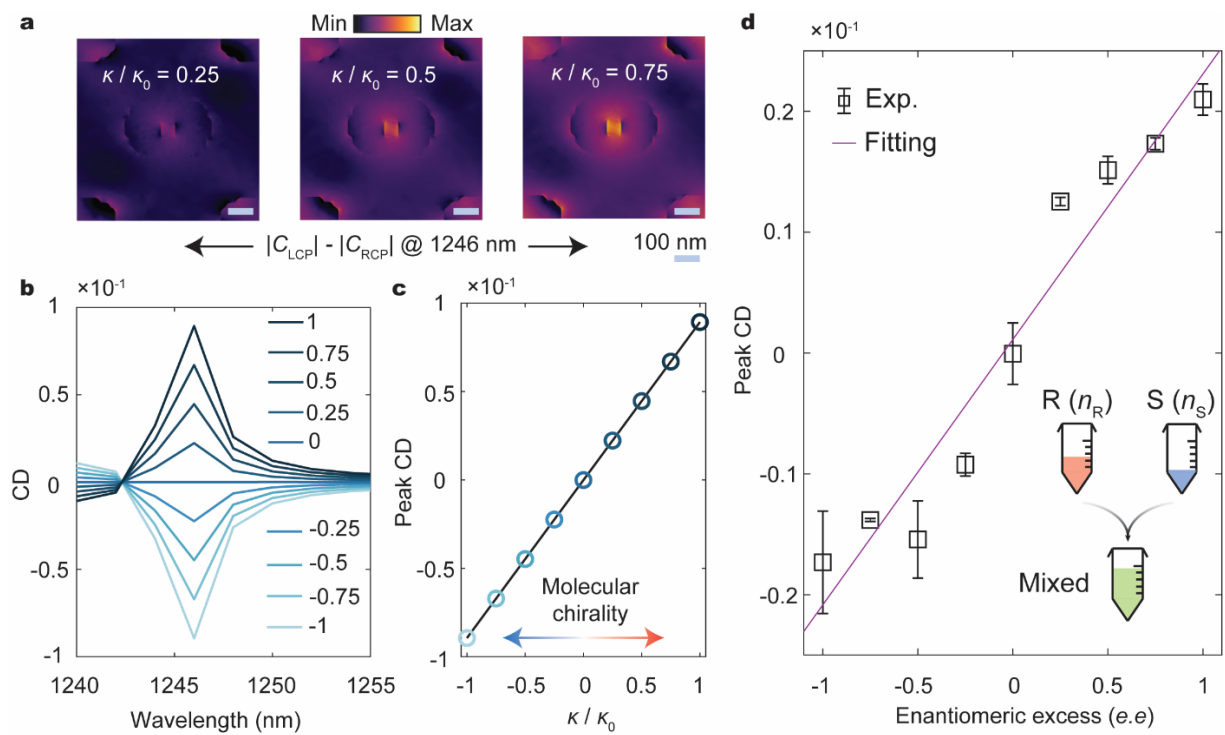

**Fig. 4 | Chiral sensor for enantiomeric excess and purity readout.** (a) Differential chiral density distribution $\Delta C = |C_L| - |C_R|$ for molecules at varying compositions (b) Simulated circular dichroism (CD) spectra for different values of the scaled Pasteur parameter $\kappa/\kappa_0$, where $\kappa_0 = -(2.3 - 0.5i) \times 10^{-3}$. (c) Peak CD as a function of $\kappa/\kappa_0$. (d) Measured peak CD for different enantiomeric compositions of (R)- and (S)-1,2-propanediol.